\documentclass[10pt]{article}
\usepackage{graphicx}

\def\Title#1{\begin{center} {\Large #1 } \end{center}}
\def\Author#1{\begin{center}{ \sc #1} \end{center}}
\def\Address#1{\begin{center}{ \it #1} \end{center}}

\newcommand\pubblock{\rightline{\begin{tabular}{l} Proceedings of the Second Annual LHCP\\ \pubnumber\\
         \pubdate  \end{tabular}}}

\newenvironment{Abstract}{\begin{quotation} \begin{center} 
             \large ABSTRACT \end{center}\bigskip 
      \begin{center}\begin{large}}{\end{large}\end{center} \end{quotation}}

\newenvironment{Presented}{\begin{quotation} \begin{center} 
             PRESENTED AT\end{center}\bigskip 
      \begin{center}\begin{large}}{\end{large}\end{center} \end{quotation}}

\def\beq{\begin{equation}}
\def\eeq#1{\label{#1}\end{equation}}
\def\eeqn{\end{equation}}

\def\beqa{\begin{eqnarray}}
\def\eeqa#1{\label{#1}\end{eqnarray}}
\def\eeqan{\end{eqnarray}}

\let\bar=\overbar

\def\Dslash{\not{\hbox{\kern-4pt $D$}}}
\def\dslash{\not{\hbox{\kern-2pt $\del$}}}

\def\msb{{\bar{\ssstyle M \kern -1pt S}}}

\usepackage{color}

\newcommand\pubnumber{ CMS CR-2014/194 }

\newcommand\pubdate{\today}

\def\affiliation{
On behalf of the CMS Collaboration, \\
Department of Physics \\
Kansas State University, Manhattan, KS 66506, U.S.A }

\begin{document}

\large
\begin{titlepage}
\pubblock

\vfill
\Title{  Searches for Leptoquarks, Extra Dimensions, and Dark Matter  }
\vfill

\Author{ LOVEDEEP KAUR SAINI }
\Address{\affiliation}
\vfill
\begin{Abstract}

We present results from several searches for various exotic physics 
phenomena like large extra
dimensions, leptoquarks, and dark matter, in proton-proton collisions at
$\sqrt{s}$ = 7 and 8~TeV delivered by the LHC and collected 
with the CMS detector. Many different
final states are analyzed using data collected in 2011 and 2012 corresponding to an integrated
luminosity up to 5.0~\textrm{fb}$^{-1}$ and 19.7~\textrm{fb}$^{-1}$. 
No sign of physics beyond the standard model has been observed so far and the results are used to set new limits on various new physics model parameters.

\end{Abstract}
\vfill

\begin{Presented}
The Second Annual Conference\\
 on Large Hadron Collider Physics \\
Columbia University, New York, U.S.A \\ 
June 2-7, 2014
\end{Presented}
\vfill
\end{titlepage}
\def\thefootnote{\fnsymbol{footnote}}
\setcounter{footnote}{0}
%

\normalsize 


\section{Introduction}

The CMS detector is a general purpose 
detector that allows one to search for 
signs of physics beyond the standard model (SM) 
at the LHC energy frontier~\cite{cmsdet}.
Among different models, the CMS collaboration searches for large 
extra dimensions, dark matter, and leptoquarks.
The analyses described here were performed using data recorded by the CMS detector at the LHC in 2011 and 2012.

\section{Leptoquarks}
The SM show a intriguing but ad hoc symmetry between quarks and
leptons that imply a more fundamental relation between the two.
In some theories beyond the SM, such as grand unification, compositeness models, and others, the existence of a new symmetry relates the quarks and
leptons in a fundamental way. These models predict the existence of new bosons,
called leptoquarks that carry both baryon and lepton numbers. The leptoquark (LQ) has fractional electric charge,
and decays to a charged lepton and a quark with unknown branching fraction $\beta$, or a neutrino and a quark with branching fraction $(1-\beta$).
 
\begin{figure}[htb]
\centering
\includegraphics[height=1.9in]{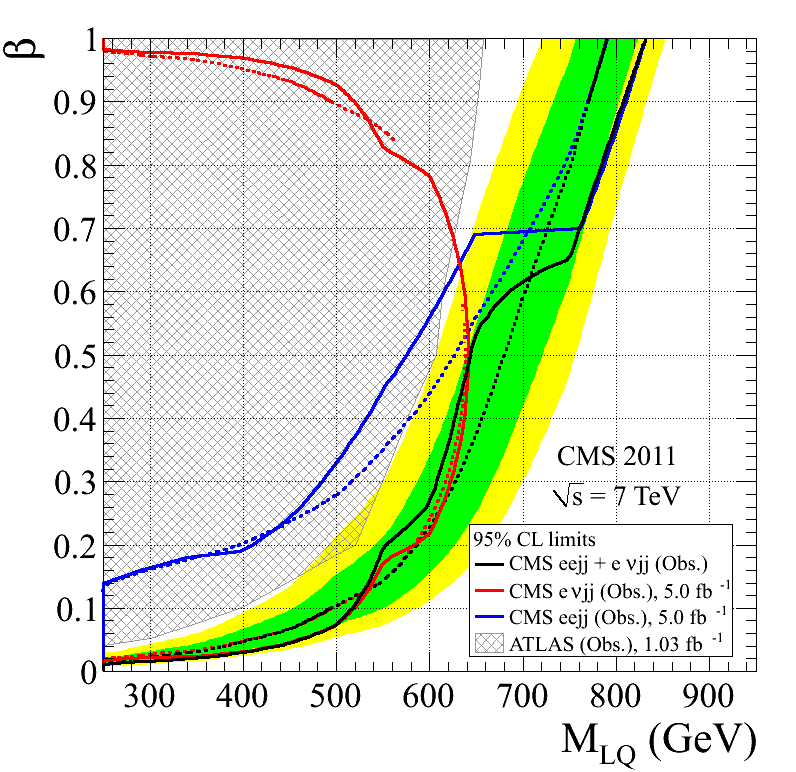}
\includegraphics[height=2in]{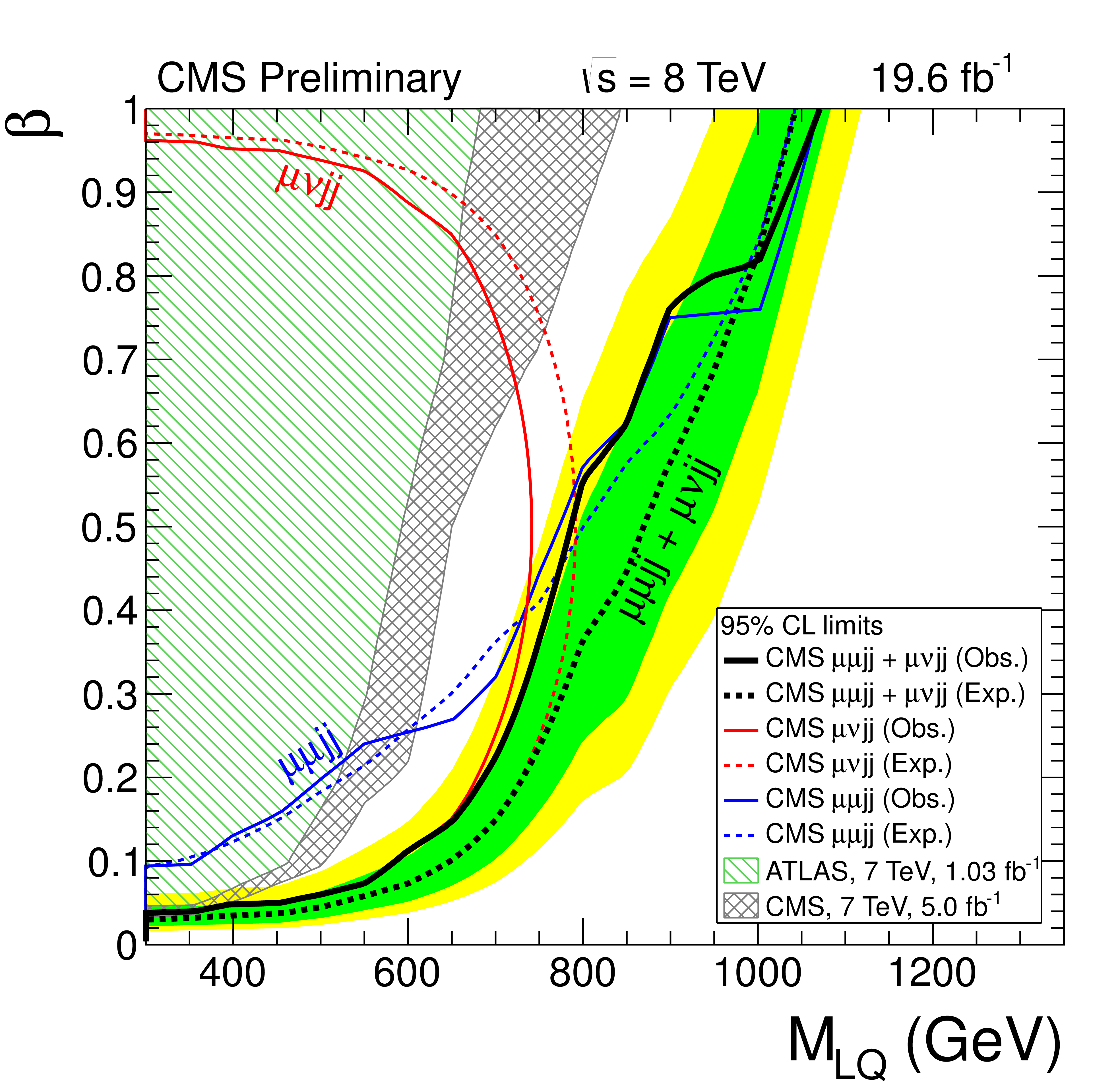}
\caption{ The expected and observed
exclusion limits at 95\% CL on the mass of first and second generation LQ as a function of the 
branching fraction $\beta$.}
\label{lq12fig}
\end{figure}

Searches for pair-production of scalar LQs for the first and second generation have been
performed in the eejj, e$\nu$jj, $\mu\mu$jj, and $\mu\nu$jj final states~\cite{lq11,lq22} by CMS collaboration. 
The scalar sum of transverse momentum ($S_\textrm{T}$) of leptons and jets (and missing transverse energy, MET, for $\ell\nu$jj)
is studied as the sensitive variable. 
The most stringent lower limits from these studies on the mass of first and second generation LQs are set to 
830~(640)~GeV and 1070~(785)~GeV, respectively, for $\beta$ = 1 (0.5) (Fig.~\ref{lq12fig}).

\begin{figure}[htb]
\centering
\includegraphics[height=1.9in]{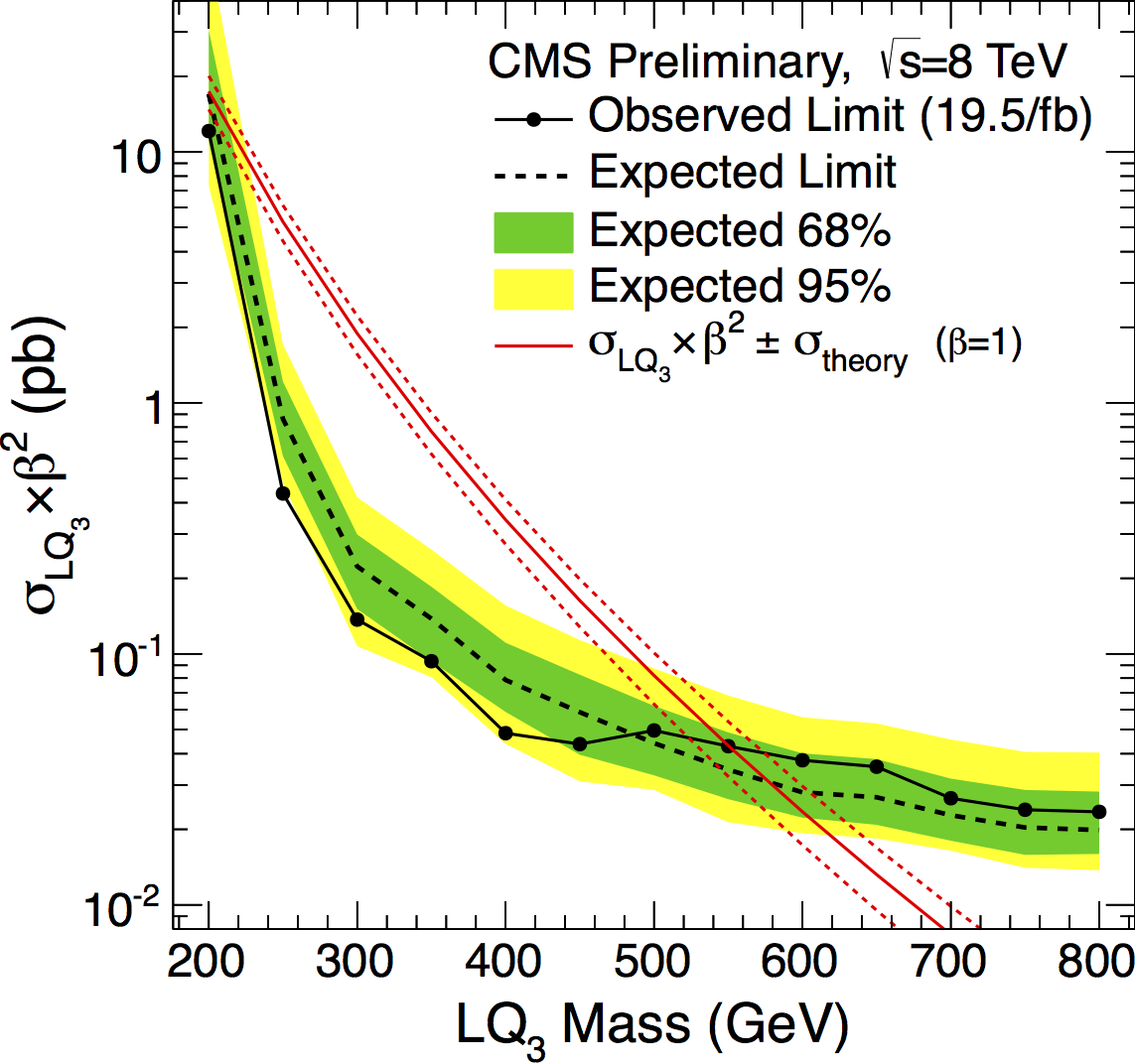}
\includegraphics[height=2in]{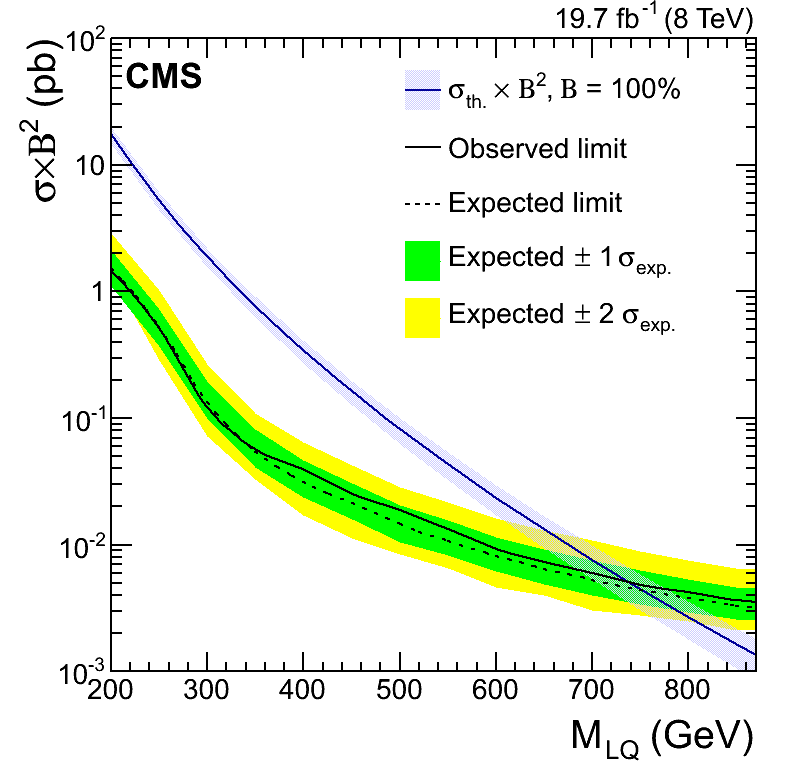}
\caption{ The expected and observed
combined upper limits on the third-generation LQ pair production 
$\sigma \times \beta^{2}$ in $t+\tau$ channel (left) and $b+\tau$ 
channel (right) at 95\% CL, as a function of the LQ mass.}
\label{lq3fig}
\end{figure}

Searches for pair-production of third generation leptoquarks have been per-
formed by the CMS experiment in both the $t+\tau$~\cite{LQ31} and 
$b+\tau$~\cite{LQ32} channels. The $t+\tau$ channel focuses on a signature
with a same-sign pair of $\mu$ and hadronically-decaying $\tau_{had}$, accompanied by
two or more hadronic jets and with a high $S_T >$ 400~GeV. 
The $b+\tau$ channel considers both $e+\tau_{had}$ and $\mu+\tau_{had}$ signatures,
accompanied by two or more hadronic jets of which at least one is b-tagged.  
Lower limits on the mass of the third generation leptoquarks are set to 
550~GeV ($t+\tau$ channel) and 740~GeV ($b+\tau$ channel) as shown in Fig.~\ref{lq3fig}.

\section{Dark Matter Searches}

\begin{figure}[htb]
\centering
\includegraphics[height=2in]{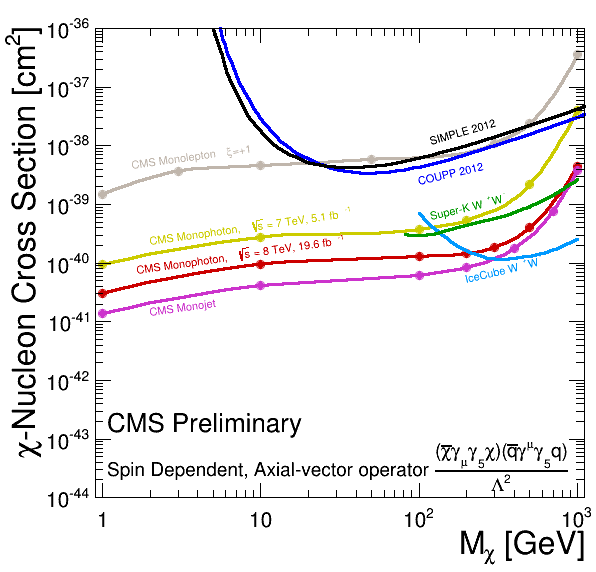}
\includegraphics[height=2in]{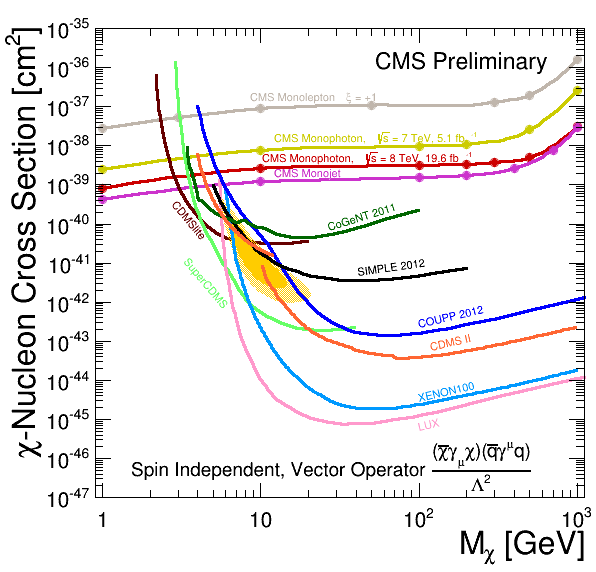}
\caption{ Limits on
spin-dependent (left) and spin-independent (right) WIMP-nucleon cross section
as a function of WIMP mass from CMS monojet and monophoton searches compared to direct and
indirect detection experiments.}
\label{DMfig}
\end{figure}

At the LHC, dark matter (DM) production can be described by an effective
field theory (EFT), assuming a contact interaction between SM and DM particles.
The characterizing parameters are the scale of
the effective interaction, $\Lambda$, and mass of the dark matter candidate, $M_{\chi}$. 
In pp collisions, DM particles may be produced in pairs 
and cannot directly be detected. 
However, if produced with a initial state radiation jet or 
a photon, the final state signature would consist of a 
single jet (monojet) or a photon (monophoton) and an inbalance of the transverse energy (MET). 
The same experimental signature is common from processes involving
 neutrinos or exotic weakly interacting particles (WIMPs).
Using the EFT, the collider results can be used to derive limits on the WIMP-
nucleon cross section as a function of the mass of the dark matter candidate as
shown in Fig.~\ref{DMfig} for the monojet~\cite{DM1} and monophoton~\cite{DM2} studies at CMS. 
The CMS limits are more stringent than the ones from direct and
indirect detection experiments for the spin-dependent WIMP-nucleon scattering
over the entire WIMP mass ($M_{\chi}$) range. For the spin-independent scattering, the
collider limits are the most stringent ones for $M_{\chi} <$ 10~GeV. 

\section{Large Extra Dimensions Searches}
\begin{figure}[htb]
\centering
\includegraphics[height=2in]{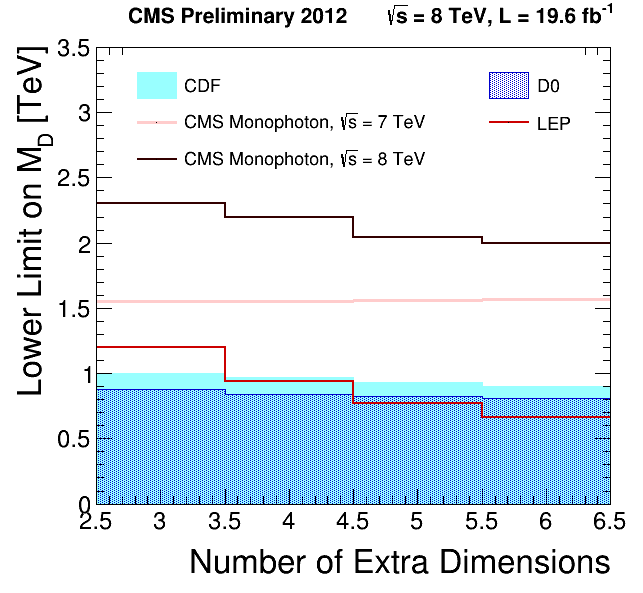}
\includegraphics[height=2in]{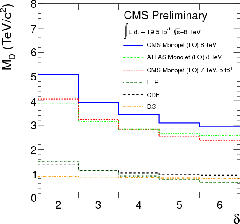}
\includegraphics[height=2in]{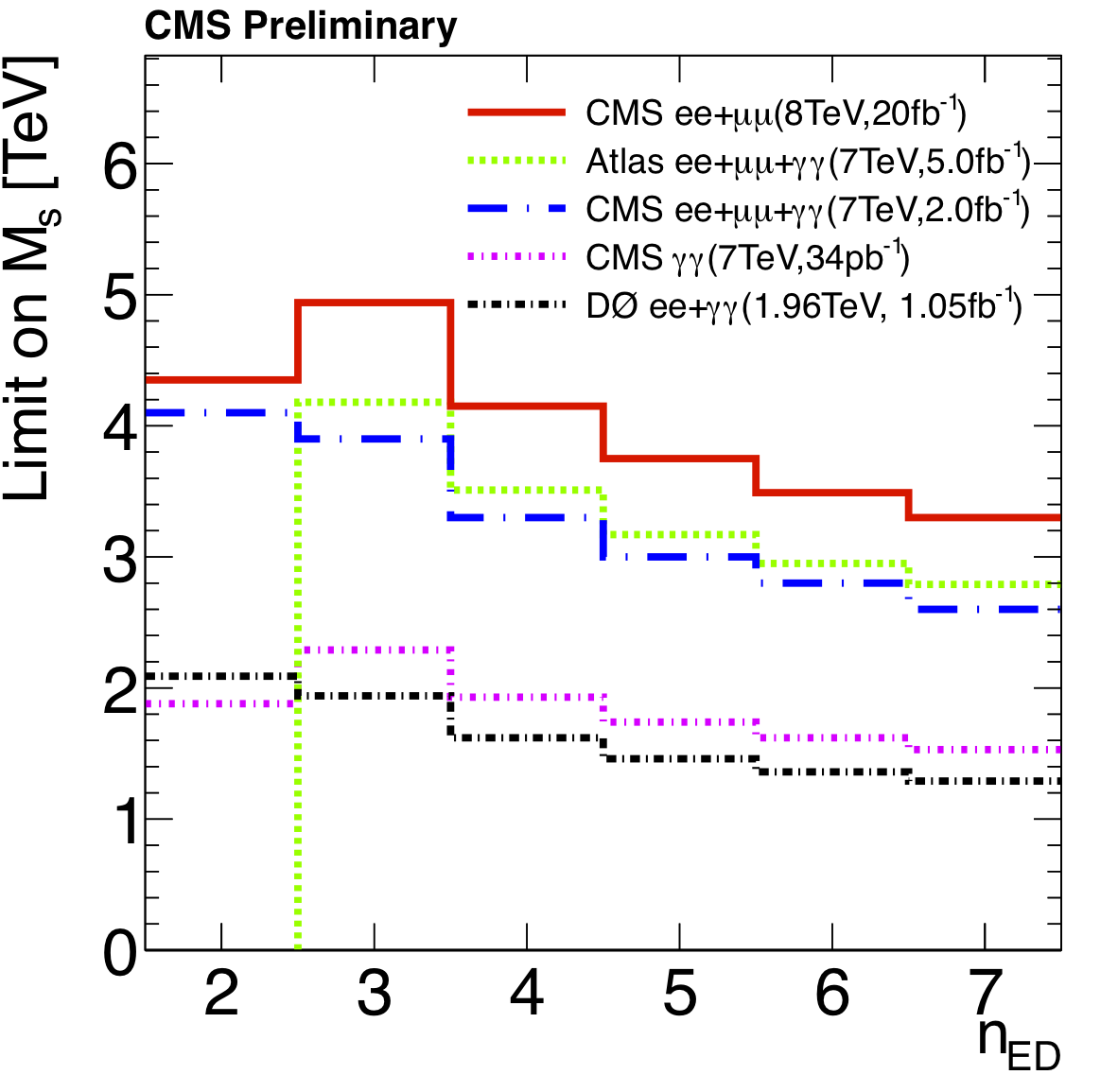}
\caption{ Limits on
the effective scale ($M_D$) as a function of extra dimensions from mono-photon (left), mono-jet (middle), and di-lepton (right) analyses, compared to results from similar searches at the Tevatron and LEP along with the CMS 7~TeV results.}
\label{LEDfig}
\end{figure}

Large extra dimensions have been proposed as a 
possible solution to the hierarchy
problem in the SM. 
The ADD model proposed by Arkani-Hamed, 
Dimopoulos, and Dvali, is postulated to have $n-$extra dimensions 
that are compactified over a multidimensional torus with radii R. 
Gravity is free to propagate into the extra dimensions, 
while SM particles and interactions are confined to ordinary space-time.
 The observed weakness of gravity 
is explained as a consequence
of the universe having ``large extra dimensions'', 
where gravity could propagate. The
number of extra dimensions ($n_D$) and 
the effective scale ($M_D$) are the main parameters of the
ADD model. Since gravitons are free to propagate in the extra dimensions, they escape the detector and can only be inferred from MET signal in the detector. 
When produced in association with a jet or a $\gamma$, 
this gives rise to the monojet or monophoton final state. 
The CMS experiment has performed searches for a production of a graviton in an association with either a jet or a photon~\cite{DM1, DM2} and limits are set on the
effective scale $M_D$ in the ADD model.
Searches for a possible enhancement of di-lepton events at high
invariant masses due to virtual graviton processes in the ADD model are also done at CMS~\cite{LEDdie, LEDdimu}.
The limits from mono-photon, mono-jet and di-lepton studies
are shown in Fig.~\ref{LEDfig}. 

\section{Conclusions}
The results obtained so far using 19.7~\textrm{fb}$^{-1}$ of pp data collected by the CMS detector
at LHC are found to be consistent with the SM predictions. These
studies set the stringent limits on various models 
of physics beyond the SM.
The search for exotic
physics at CMS is still a work in progress, and a few more studies using 
the full 2012 CMS data are going on. But we eagerly wait for the era of
$\sqrt{s}$ = 13~TeV collisions at LHC, where we will have significantly greater
sensitivity on account of both the higher luminosity and center of mass
energy.
%
%
%


\end{document}